\pdfoutput=1
\documentclass[prd,aps,superscriptaddress,amsmath,amssymb,twocolumn,nobibnotes,nofootinbib,10pt,groupedaddress]{revtex4-2}

\usepackage{braket,verbatim,bm,bbold,amsmath,amssymb,epsfig,xcolor,float}
\definecolor{redCB}{RGB}{150,50,60}  
\definecolor{dteal}{RGB}{7,94,84}  %teal dark green
\definecolor{blueCB}{RGB}{108,144,170}
\usepackage[colorlinks=true,    	
			pdfstartview=FitV,
            bookmarks=false,
            citecolor=dteal,
            linkcolor=dteal,
           linkcolor=dteal,
            hyperfootnotes=true,
            linktoc=page,
            urlcolor=dteal]{hyperref}
\hypersetup{breaklinks=true}
\usepackage{soul}
\usepackage{tikzlings}
\usepackage{tikzducks}
\usepackage{flushend}
\usepackage{pbalance}

\newcommand{\nc}{\newcommand}
\nc{\ir}{\mathrm{i}}
\nc{\dd}{\mathrm{d}} 
\nc{\eE}{\mathrm{e}}
\nc{\Tr}{\text{Tr}}
\nc{\id}{\mathbb{I}}
\nc{\Z}{\mathcal{Z}}
\nc{\E}{\mathcal{E}}
\nc{\F}{\mathcal{F}}
\nc{\Om}{\Omega}
\nc{\N}{\mathcal{N}}
\nc{\spp}{\hspace{1pt}}
\nc{\spm}{\hspace{-1pt}}
\nc{\lb}{\label} 
\nc{\nn}{\nonumber} 
\nc{\ra}{\rangle} 
\nc{\la}{\langle} 
\nc{\Nn}{\mathcal{N}} 
\nc{\sigb}{\boldsymbol{\sigma_{\text{bd}}}}
\nc{\Hh}{\mathcal{H}}
\nc{\HA}{\mathcal{H}_A}
\nc{\HB}{\mathcal{H}_B}
\nc{\HC}{\mathcal{H}_C}
\nc{\ta}{\tilde{a}}
\nc{\eps}{\epsilon}
\nc{\Q}{\mathcal{Q}}
\nc{\T}{\mathcal{T}}
\nc{\Oo}{\mathcal{O}}


\def\bea#1\eea{\begin{align}#1\end{align}}
\def\bes#1\ees{\begin{equation}\begin{split}#1\end{split}\end{equation}}

\makeatletter                                                                 
    \def\footnoterule{\kern -6\p@        % to increase vertical space between rule and notes ; difference between the values after "kern" is the width of the rule!
  \hrule \@width 0.5in \kern 5.7\p@}  % the in value is the length of the footnoterule
\makeatother

\begin{document}

%%%%%%%%%%%%%%%%%%%%%%%%%%%%%%%%%%%%%%%%%%%%%%%%%%%%%%%%%%%%%%%%%%%
\title{Tripartite entanglement dynamics following a quantum quench}
%\title{Tripartite entanglement dynamics}
%%%%%%%%%%%%%%%%%%%%%%%%%%%%%%%%%%%%%%%%%%%%%%%%%%%%%%%%%%%%%%%%%%%

\author{Cl\'ement Berthiere}
\email{clement.berthiere@irsamc.ups-tlse.fr}
\affiliation{\vspace{3pt}Laboratoire de Physique Théorique, CNRS, Université de Toulouse, France}

\date{\today}

\begin{abstract}
We investigate the dynamics of multipartite entanglement after quenches from initial states which generate multiplets of quasiparticle excitations beyond the usual pair structure. 
We focus on the dynamics of tripartite entanglement through the lens of the Markov gap---a computable quantity that signals irreducible tripartite entanglement when positive.
In the tight-binding chain, we show that the Markov gap is positive at intermediate times, implying the presence of tripartite entanglement. After a time delay, the Markov gap  increases and then decays at longer times, thus exhibiting an entanglement barrier. We argue that those qualitative features are consistent with an interpretation of the spreading of tripartite entanglement by triplets of tripartite-entangled quasiparticles.
\end{abstract}

\maketitle
\makeatletter

\def\l@subsubsection#1#2{}
\makeatother

%\tableofcontents

The investigation of non-equilibrium dynamics in isolated many-body quantum systems has emerged as a central topic in modern physics. Understanding how entanglement and correlations spread in these systems is not only fascinating, but also of fundamental importance across various domains, including condensed matter, statistical physics, and quantum field theory. One of the most studied protocols is the quantum quench (e.g., \mbox{\cite{Calabrese:2007rg,Polkovnikov:2010yn,Gogolin:2015gts,Calabrese:2016psf,Essler:2016ufo})} in which an extended system evolves unitarily under an Hamiltonian after having being initially prepared in a non-equilibrium state.
Recent progress on the experimental front \cite{2012NatPh...8..277B,Monroe:2019asq,2020NatPh..16..132B,Joshi:2021kzb} has made possible the engineering of Hamiltonian dynamics of isolated quantum systems, and the measuring of non-trivial physical properties such as the growth of entanglement following a quantum quench \cite{2016Sci...353..794K,2019Sci...364..260B,Elben:2020hpu} and out-of-time ordered correlators \cite{Li:2016xhw,Garttner:2016mqj,Landsman:2018jpm,Joshi:2020quh}.

Much of the research has focused on the time evolution of bipartite entanglement measures after a quantum quench \cite{Calabrese:2005in,DeChiara:2005wb,2008PhRvA..78a0306F,2008JSMTE..05..018L,2011JSMTE..08..019S,2013PhRvL.111l7205K,2014EL....10740002K,Coser:2014gsa,Nahum:2016muy,Cotler:2016acd,2016PhRvA..93e3620B,2017ScPP....2....2D,vonKeyserlingk:2017dyr,Alba:2017qag,Hackl:2017ndi,Alba:2017lvc,Wen:2017smb,Alba:2019ybw,Parez:2020vsp,Parez:2021pgq,Alba:2021qpx,Bertini:2022fnr,Alba:2022enm,Parez:2022egh,Berthiere:2023gkx}, such as the entanglement entropy. 
Many qualitative features of the entanglement dynamics of integrable many-body quantum systems may be understood using a simple physical description put forward in \cite{Calabrese:2005in}. Within the so-called quasiparticle picture, the initial state is regarded as a source of pairs of entangled quasiparticle excitations which ballistically propagate across the system and carry entanglement. The type of entangled quasiparticles responsible for the entanglement spreading is dictated by the initial state. Recently, the focus has expanded from conventional quenches generating pairs of entangled quasiparticles to richer quenches that give rise to entangled multiplets of quasiparticles \cite{2018JSMTE..06.3104B,Bastianello:2018fvl,Roy:2021efl,Caceffo:2023hns}, which are particularly relevant in the context of multipartite entanglement dynamics.

The structure of entanglement becomes significantly more complex when more than two parties are involved \cite{horodecki2009quantum,guhne2009entanglement}, leading to the emergence of several inequivalent classes of entanglement. To comprehensively understand the entanglement structure of a quantum system, it is essential to investigate its multipartite entanglement properties, see, e.g., \cite{Akers:2019gcv,Zou:2020bly,Hayden:2021gno,Liu:2021ctk,tam2022topological,Parez:2022ind,Liu:2023pdz,Berthiere:2023bwn,Parez:2024zbz}. Recent theoretical activities have explored the time evolution of the tripartite mutual information after a quantum quench, e.g., \cite{Iyoda:2017pxe,Schnaack:2018bhs,Carollo:2022lrq,Maric:2022rsc,Caceffo:2023hns}. In out-of-equilibrium systems, a negative tripartite mutual information is used as a diagnostic of scrambling \mbox{\cite{Hayden:2007cs,Sekino:2008he,Hosur:2015ylk},} associated with the notion of chaos.
Importantly, it should be noted that the tripartite mutual information is \textit{not} a measure of three-party entanglement. \mbox{Tripartite} entanglement, representing the simplest level of complexity beyond bipartite entanglement, deserves to be explored with particular attention.

In this letter, we investigate the dynamics of tripartite entanglement following a quantum quench producing triplets of quasiparticles. To that aim, we consider the Markov gap, a computable quantity that signals irreducible tripartite entanglement when positive \cite{Zou:2020bly}. We show that this quantity is positive at intermediate times in the hydrodynamic regime, implying the presence of tripartite entanglement. The time evolution of the Markov gap is consistent with an interpretation of the spreading of tripartite entanglement by triplets of tripartite-entangled quasiparticles. Conversely, the Markov gap can be exploited to witness the existence of multiplets with at least three (entangled) excitations.

\begin{figure}[t]
%\vspace{.2cm}
\centering
\includegraphics[scale=0.94]{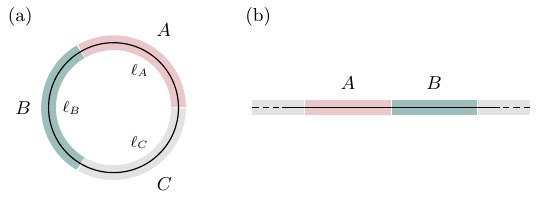}
%\vspace{-15pt}
\caption{(a) A finite system divided into three parts $A,\spp B$ and $C$, of respective size $\ell_A,\spp \ell_B$ and $\ell_C$. (b) An infinite system ($\ell_C\rightarrow\infty$) with adjacent subsystems $A$ and $B$.}
\lb{fig_tri}
\end{figure}

\begin{figure*}
%\vspace{.2cm}
\centering
\includegraphics[scale=1]{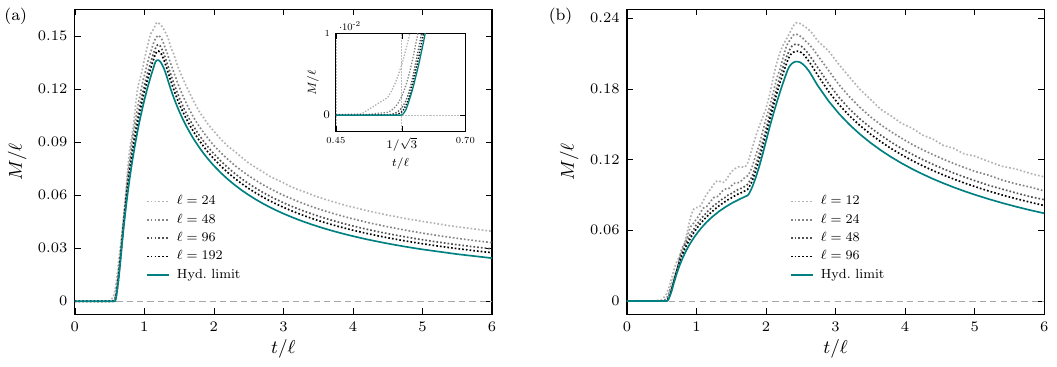}
\vspace{-15pt}
\caption{Dynamics of the Markov gap in the tight-binding chain after a quench from \eqref{eq:state}. The dotted lines correspond to different subsystem lengths $\ell=\min(\ell_A,\ell_B)$, while the solid teal lines are the extrapolations to the hydrodynamic limit \mbox{$t,\ell\rightarrow\infty$} with $t/\ell$ fixed. 
(a) Equal subsystem sizes. The inset shows a close-up around time \mbox{$t=\ell/\sqrt3$} before which the Markov gap is zero.
(b) Unequal subsystem sizes $\ell_B=3\ell_A$. A time delay $t_d=\ell/\sqrt3$ and a secondary increase at time $t_b=\ell\sqrt3$ are observed.}
\lb{fig2}
\end{figure*}

\medskip
\textit{Markov gap} --- We first review the definitions of the mutual information $I(A:B)$ and the reflected entropy $S_R(A:B)$, from which is defined the Markov gap.

The mutual information is a measure of correlation between two subsystems of a general quantum state. It is defined as
$I(A:B) = S(A) + S(B) - S(A\cup B)$,
where $S(X)=-\Tr\spp(\rho_X\spm\log\rho_X)$ is the entanglement entropy of $\rho_X$ on subsystem $X$.

For the reflected entropy \cite{Dutta:2019gen} (see also, e.g., \cite{Jeong:2019xdr,Kudler-Flam:2020url,Bueno:2020vnx,Li:2020ceg,Berthiere:2020ihq,Bueno:2020fle,Akers:2022max,Vasli:2022kfu,Lu:2022cgq,Afrasiar:2022fid,Sohal:2023hst}), we consider the canonical purification $\ket{\sqrt{\rho_{AB}}}$ of $\rho_{AB}$, where we regard the operator $\sqrt{\rho_{AB}}$ as a state in a doubled Hilbert space $\Hh_A\otimes \Hh_B\otimes\Hh_{A^*}\spm\otimes \Hh_{B^*}$. The reflected entropy is then defined as $S_R(A:B)=S(AA^*)$, i.e. the entanglement entropy associated to $\ket{\sqrt{\rho_{AB}}}$. 

The Markov gap $M$ is the difference between reflected entropy and mutual information: $M \equiv S_R-I$. The reflected entropy being lower bounded by the mutual information \cite{Dutta:2019gen}, the Markov gap is a nonnegative quantity.
For Gaussian states, these quantities can be computed using the two-point correlation function \cite{peschel2003calculation,peschel2009reduced,Bueno:2020vnx,Berthiere:2023gkx}, as explained in Sec.\,\ref{Apdx:gaussian} of Supplemental Material (SM) \cite{suppmat}.

For pure tripartite states, a positive Markov gap \mbox{$M\spm>\spm0$} signals tripartite entanglement that cannot be reduced to bipartite entanglement with local unitaries \cite{Zou:2020bly} (see also \cite{Akers:2019gcv,Hayden:2021gno,Siva:2021cgo,Berthiere:2023bwn} for further studies). %,
%\bea
%M>0 \quad\Rightarrow\quad \text{tripartite entanglement}\spp.
%\eea
Note that the Markov gap does not detect GHZ-type entanglement, e.g.~it is zero for the GHZ state of three qubits.
Our objective is to study the dynamics of (non-GHZ) tripartite entanglement following quantum quenches from low-entangled initial states, through the lens of the Markov gap.

\medskip
\textit{Quench protocol and setup} --- 
A fermionic chain is prepared at $t=0$ in the state $\ket{\psi_0}$. At time $t>0$, we let the initial state evolve as $\ket{\psi(t)}=e^{-itH}\ket{\psi_0}$ under the tight-binding Hamiltonian \mbox{$H=-\frac12\sum_i(c^\dagger_i c_{i+1} + c^\dagger_{i+1} c_i)$,}
with $c_i$ standard fermionic operators, which can be mapped onto the XX spin chain.
We study the time evolution generated by the above Hamiltonian on the initial state:
\bea\lb{eq:state}
|\psi_0\ra = \prod_{i=1}^{L/3}\spm c_{3i}^\dagger|0\ra \spp,
\eea
where $L$ is the total size of the chain such that $L/3$ is integer.
This state is Gaussian and invariant under translation of three sites, and presents a single fermion per unit cell. 
In SM Sec.\,\ref{Apdx:correl}, we consider generalizations of this state to arbitrary size of unit cell and arbitrary number of fermions in each cell, as well as other states.

The state $|\psi_0\ra$ in \eqref{eq:state} generates triplets of quasiparticle excitations after the quench (e.g., \cite{2018JSMTE..06.3104B}), instead of the more conventional pair structure that leads to a growth of bipartite entanglement only as pairs of entangled quasiparticles spread in the system. It is natural to wonder whether the spreading of multiplets of quasiparticles may lead to a growth of multipartite entanglement---in the case at hand of tripartite entanglement. 

We study the time evolution of the Markov gap as a witness of tripartite entanglement after quantum quenches described above. The system is divided into three contiguous parts $A$, $B$ and $C$, see Fig.\,\hyperref[fig_tri]{\ref{fig_tri}(a)}.
For simplicity, we work in the thermodynamic limit $L\rightarrow\infty$, which we access by letting the length of subsystem $C$ be infinite (see Fig.\,\hyperref[fig_tri]{\ref{fig_tri}(b)}). We choose the characteristic length scale to be $\ell=\min(\ell_A,\ell_B)$.

\medskip
\textit{Tripartite entanglement spreading} --- We study the time evolution of the Markov gap $M$ in the tight-binding chain after a quantum quench from the crystal state \eqref{eq:state}, which
generates triplets of quasiparticles. Our numerical results are reported in Fig.\,\ref{fig2} which shows data for $M/\ell$ plotted as a function of $t/\ell$. We consider two cases: equal subsystem sizes $\ell_A=\ell_B$ displayed in Fig.\,\hyperref[fig2]{\ref{fig2}(a)}, and different subsystem sizes $\ell_B=3\ell_A$ showed in Fig.\,\hyperref[fig2]{\ref{fig2}(b)}.

We find that the Markov gap is nonvanishing in the \mbox{hydrodynamic} regime \mbox{$t,\ell\rightarrow\infty$} with $t/\ell$ fixed. The hydrodynamic extrapolations (solid lines in Fig.\,\ref{fig2}) are fits %to $M_{\rm Hyd}+a/\ell + b/\ell^2$, with $a, b$ fitting parameters, 
of our finite-size results for $90\le\ell\le2070$.
We observe a time delay $t_d=\min(\ell_A,\ell_B)/\sqrt3$ \mbox{before} which the Markov gap is zero. At time \mbox{$t>t_d$,} the Markov gap becomes positive and increases rapidly, \mbox{signaling} a production of tripartite entanglement. A second sharp increase is observed at time $t_b=\max(\ell_A,\ell_B)/\sqrt3$ for $\ell_A\neq\ell_B$, see Fig.\,\hyperref[fig2]{\ref{fig2}(b)} (note that $t_b=t_d$ for $\ell_A=\ell_B$). A plateau develops as one of the subsystems becomes larger than the other. After reaching its maximum value, the Markov gap decreases and vanishes in the large-time limit, thus displaying an entanglement barrier. The decay is slow, as $M\sim t^{-1}$, which means that tripartite entanglement persists in the system at all finite times $t>t_d$. 

The features described above are consistent with a quasiparticle interpretation of the spreading of tripartite entanglement. The initial state acts as a source of triplets of quasiparticles produced homogeneously throughout the system at time $t=0$.
Quasiparticles emitted from the same position are assumed to be entangled. As they propagate ballistically for $t>0$, quasiparticles carry and spread entanglement throughout the system.
Entanglement between the three parties $A, B, C$ at time $t$ is thus related to the triplets of entangled quasiparticles that are shared between $A, B, C$, see Fig.\,\ref{fig_qpp}. 

Initial states that produce only pairs of quasiparticles, such as the Néel and dimer states, cannot entangle more than two parties and are thus expected to yield a nonextensive Markov gap in the hydrodynamic regime. We have checked that this is indeed the case. For initial states that give rise to multiplets of quasiparticle excitations, we expect that multipartite entanglement can be accordingly produced after a quench.

\medskip
\textit{Discussion} --- 
Multipartite entanglement is difficult to grasp in quantum many-body systems, even more so in out-of-equilibrium settings. Through the lens of the Markov gap, we have investigated the dynamics of tripartite entanglement following a quantum quench in the XX chain from an initial state generating triplets of quasiparticle excitations. The Markov gap, when positive, signals irreducible tripartite entanglement. We have shown that this quantity is positive at intermediate times in the hydrodynamic regime, implying the presence of tripartite entanglement.  
After a time delay, the Markov gap increases and then decays at longer times, thus exhibiting an entanglement barrier. Those qualitative features are consistent with an interpretation of the spreading of tripartite entanglement by triplets of tripartite-entangled
\begin{figure}[H]
%\vspace{.2cm}
\centering
\includegraphics[scale=1.25]{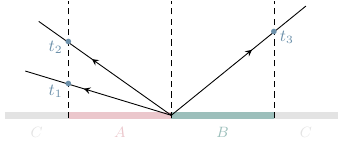}
\vspace{-8pt}
\caption{Within the quasiparticle picture, only multiplets that are shared between all the subsystems $A$, $B$, $C$ contribute to the Markov gap. Here is depicted a triplet produced at the boundary of $A$ and $B$. The times $t_1<t_2<t_3$ at which quasiparticles change subsystem are shown. For $0\le t \le t_1$, the triplet is shared between $A$ and $B$ only; the Markov gap is thus zero. At $t=t_1$, the leftmost quasiparticle leaves $A$ to enter $C$, while the second and third quasiparticles stay in $A$ and $B$, respectively, hence the triplet starts to contribute to the Markov gap. For $t>t_2$, the two left quasiparticles have left $A$ such that the Markov gap vanishes again.}
\lb{fig_qpp}
\end{figure}
\hspace*{-10pt}quasiparticles. We expect that $n$-partite entanglement can be similarly produced after quantum quenches from initial states generating $n$-plets of quasiparticles.

We have performed numerical calculations to access the hydrodynamic regime of large subsystems and long times, with their ratio fixed. 
The entanglement entropy following a quench from the initial state \eqref{eq:state} for a single interval of length $\ell$ can be obtained analytically and reads
\bea
S(t) %&= s(1/3)\spm\spm\int_{-\pi}^{\pi}\frac{dk}{2\pi}\min\spm\big(\ell,t|v(k)-v(k-2\pi/3)|\big),\\
&=s(1/3)\spm\spm\int_{-\pi}^{\pi}\frac{dk}{2\pi}\min\spm\big(\spm\sqrt3\spp|v_k| \hspace{0.35pt} t,\ell\big)\spp,
\eea
where $s(x)=-x\log x-(1-x)\log(1-x)$ and $v_k=\sin k$. The mutual information of two adjacent subsystems can easily be inferred from this formula. It would be interesting to analytically compute the reflected entropy, to obtain the full analytical form of the Markov gap. %We leave this as future direction.

%\balance
%\vspace*{5pt}
\medskip
\textit{Acknowledgments} --- 
It is a pleasure to thank Vincenzo Alba and Yijian Zou for enlightening correspondences. I am grateful to Gilles Parez for many interesting discussions and valuable comments on the draft.

%\clearpage

%\pagebreak

\let\oldaddcontentsline\addcontentsline% Store \addcontentsline
\renewcommand{\addcontentsline}[3]{}% Make \addcontentsline a no-op
\bibliographystyle{utphys} 
%\bibliography{}
\bibliography{tripartite_dynamics.bib}

\let\addcontentsline\oldaddcontentsline% Restore \addcontentsline

%%%%%%%%%% Merge with supplemental materials %%%%%%%%%%
\onecolumngrid
\clearpage
\begin{center}
\hspace*{-5pt}\mbox{\textbf{\large Supplemental Material: Tripartite entanglement dynamics following a quantum quench}}\\[.45cm]
  Cl\'ement Berthiere\\[.15cm]
  {\itshape \small 
  Laboratoire de Physique Théorique, CNRS, Université de Toulouse, France\\}
{\small (Dated: \today)}%\vspace*{-0.55cm}
\end{center}

%%%%%%%%%% Merge with supplemental materials %%%%%%%%%%
%%%%%%%%%% Prefix a "S" to all equations, figures, tables and reset the counter %%%%%%%%%%
\setcounter{equation}{0}
\setcounter{figure}{0}
\setcounter{table}{0}
\setcounter{page}{1}
\makeatletter
\renewcommand{\thesection}{\Roman{section}}
\renewcommand{\thesubsection}{\arabic{subsection}}
\renewcommand{\theequation}{S\arabic{equation}}
\renewcommand{\thefigure}{S\arabic{figure}}
\renewcommand{\thetable}{S\Roman{table}}
%\renewcommand{\bibnumfmt}[1]{[S#1]}
%\renewcommand{\citenumfont}[1]{S#1}
%%%%%%%%%% Prefix a "S" to all equations, figures, tables and reset the counter %%%%%%%%%%

\begin{comment}
\vspace{-10pt}
\def\l@subsection#1#2{}
\addtocontents{toc}{\vspace{-10pt}}
\renewcommand{\baselinestretch}{0.75}\normalsize
\tableofcontents % works great.  Includes supplement in toc, as it should.
\renewcommand{\baselinestretch}{1.0}\normalsize
\end{comment}

\medskip

%\begin{comment}

\setcounter{section}{0}

\section{Markov gap for Gaussian fermionic states}
\lb{Apdx:gaussian}

For Gaussian fermionic states with conserved particle number, the entanglement entropy of a subsystem $A$ can be computed using the standard formula \cite{peschel2003calculation,peschel2009reduced}
\bea
S(A)  = -\Tr\big[C_A \log C_A + (\mathbb{1}-C_A)\log(\mathbb{1}-C_A)\big],
\eea
where $C$ is the two-point correlation matrix with elements
$C_{ij}=\Tr(\rho c^\dagger_i c_j)$, and $C_A$ is its restriction to subsystem $A$.

The reflected entropy can be computed similarly \cite{Bueno:2020vnx,Berthiere:2023gkx} in terms of the ``reflected" correlation matrix 
\bea\lb{correl_FF}
{\bf C}=\spm
\begin{pmatrix}
C_{AB} & \sqrt{C_{AB}(\mathbb{1}-C_{AB})}\\
\sqrt{C_{AB}(\mathbb{1}-C_{AB})} & \mathbb{1}-C_{AB}
\end{pmatrix}\spm\spm,
\eea
where $C_{AB}$ is the correlation matrix for $\rho_{AB}$. Introducing ${\bf C}_A$ as the restriction of ${\bf C}$ to subsystem $A$, the reflected entropy is obtained as
\bea
S_R(A:B) = -\Tr\big[{\bf C}_A \log {\bf C}_A + (\mathbb{1}-{\bf C}_A)\log(\mathbb{1}-{\bf C}_A)\big].
\eea
With these formulas in hand, one can easily compute the Markov gap defined in the main text.

\section{Time-dependent correlation functions in the thermodynamic limit}
\lb{Apdx:correl}

We consider quenches to the tight-binding model with Hamiltonian 
\bea
H = -\frac12\sum_j^L\Big(c^\dagger_j c_{j+1} + c^\dagger_{j+1} c_j \Big),
\eea
on a periodic chain of length $L$. The $c_j$'s are the fermionic annihilation operators at site $j$. The Jordan-Wigner transformation maps the model onto the XX spin chain.
The tight-binding Hamiltonian can be put in a diagonal form
\bea
H=\sum_k\eps_k d_k^\dagger d_k\spp,\qquad {\rm where}\qquad 
\eps_k=-\cos k\spp, \quad d_k=\frac{1}{\sqrt L}\sum_je^{-ijk}c_j\spp, \quad k=\frac{2\pi m}L,\;m=1,2,\cdots,L\spp.
\eea

Since all the states we consider are Gaussian, the time evolution after the quench is completely characterized by the two-point correlation matrix
\bea\lb{correl_def}
C_{ij}(t) = \la\psi_0|e^{itH} c_i^\dagger c_j e^{-itH}|\psi_0\ra\spp.
\eea
The fermion operator in the Heisenberg picture reads
\bea
e^{itH} c_j^\dagger e^{-itH} = \frac1L \sum_{k,l}e^{-i(j-l)k}e^{it\eps_k} c_l^\dagger\spp.
\eea
Plugging this into \eqref{correl_def} yields
\bea
\lb{correl_gen}
C_{ij}(t) = \frac{1}{L^2} \sum_{\substack{k,k' \\ l,l'}}e^{-i(i-l)k}e^{i(j-l')k'}e^{it(\eps_k-\eps_{k'})}C_{ll'}(0)\spp.
\eea

We focus on two classes of initial states with periodic patterns. Both are invariant under translation of $p$ sites and are Gaussian. The \textit{crystal states} $|C_{p,q}\ra$ possess $q$ fermions in each unit cell of $p$ sites. In contrast, for the states $|W_p\ra$ there is in each cell a single fermion in a specific quantum state---a W state. For that reason, we call these states \textit{W-product states}.
In the following, we give explicit expression of the corresponding correlation matrices in the thermodynamic limit $L\rightarrow\infty$.

\subsection{Crystal states}
We define the class of crystal states as
\bea
|C_{p,q}\ra = \prod_{i=1}^{L/p}\spm\spm\Bigg(\prod_{j=0}^{q-1}c_{pi-j}^\dagger\Bigg)\spm|0\ra =|\underbrace{\spm\downarrow\spm\cdots\spm\downarrow}_{p-q}\underbrace{\uparrow\spm\cdots\spm\uparrow}_{q}\spm\downarrow\spm\cdots\spp\ra\spp,
\eea
where $L$ is divisible by $p$ and $0<q<p$. This state presents a periodic pattern of $p$ sites, while its filling fraction is $q/p$. For $p=2$ and $q=1$, it reduces to the well-known Néel state $|\spm\spm\downarrow\uparrow\downarrow\uparrow\spm\cdots\ra$. We have $C_{ij}(0) ={\rm diag}(\underbrace{0,\cdots,0}_{p-q},\underbrace{1,\cdots,1}_{q},\cdots)$.

In the thermodynamic limit $L\rightarrow\infty$, the time-dependent correlation matrix for a quench from crystal states reads
\bea
&C_{ij}(t) = C_{ij}(\infty) + \frac{i^{j-i}}p\spm\spm \sum_{n\scalebox{0.7}{$\in$} \omega_{p,q}}\spm\spm a_{i,j}^n\spp J_{j-i}\big(2t\sin(\pi n/p)\big)\spp,
\eea
where $J_\alpha(z)$ is Bessel's function of the first kind, and 
\bea
&C_{ij}(\infty) = \frac qp\delta_{i,j}\spp,\\
&a_{i,j}^n %&=\sum_{r=0}^{q-1}\cos\spm\Big(\frac\pi2(i - j) + \frac{\pi n}{p} (i + j + 2r)\Big),\\
 =\frac{\sin\spm\big(\spm\frac{n\pi q\spm}{p}\big)}{\sin\spm\big(\spm\frac{n\pi}{p}\spm\big)}\cos\spm\Big(\frac\pi2(i-j) + \frac{\pi n}{p} (i + j+q -1)\spm\Big),\\
&\omega_{p,q} =[1,p-1]\backslash \Big\{\frac{p}{d_{p,q}},\frac{2p}{d_{p,q}},\cdots,p-\frac{p}{d_{p,q}}\Big\},
\eea
with $d_{p,q}$ the greatest common divisor of $p$ and $q$. 

%For the Néel state with $p=2$ and $q=1$, we recover
%\bea
%C_{ij}(t) = \frac12\delta_{ij} + \frac{(-1)^j}{2i^{i-j}} J_{i-j}(2t)\spp.
%\eea

In the main text, we focus on the initial state $|\psi_0\ra\equiv|C_{3,1}\ra=|\spm\spm\downarrow\downarrow\uparrow\spm\cdots\ra$, for which the correlation matrix reads
\bea
&C_{ij}(t) = \frac 13\delta_{ij} + \frac{2(-1)^j}{3i^{i-j}}\cos\Big(\frac{\pi}{6}(i+j)\Big)\spp J_{i-j}\big(t\sqrt{3}\big)\spp.
\eea

\subsection{W-product states}
We define the class of W-product states as
\bea
|W_p\ra = \frac{1}{p^{L/2p}}\prod_{i=1}^{L/p}\spm\Bigg(\sum_{j=0}^{p-1}c_{pi-j}^\dagger\Bigg)\spm|0\ra \spp,
%=|\spm\spm\downarrow\cdots\downarrow\spp\uparrow\spp\downarrow\cdots\downarrow\spp\uparrow\spp\downarrow\cdots\ra
\eea
where $L$ is divisible by $p$, and $1/p$ is the filling fraction. For $p=2$, it reduces to the dimer state which is the product of triplets states $\frac{1}{\sqrt2}(|\spm\spm\uparrow\downarrow\ra+|\spm\spm\downarrow\uparrow\ra)$. %, i.e.
%\bea
%C_{ij}(t) = \frac12\delta_{ij} + \frac14\big(\delta_{i,j+1}+\delta_{i,j-1}\big) +  \frac{i^{i+j+1}}4\big(J_{j-i-1}(2t)+J_{j-i+1}(2t)\big)\spp.
%\eea
For $p\ge3$, the state $|W_p\ra$ is a product of W states which are representatives of non-biseparable states. We have $C(0) = \frac1p( \mathbb{1}_p \oplus\cdots\oplus\mathbb{1}_p)$, with $\mathbb{1}_p$ the unit constant matrix of size $p$.

%\pagebreak
The time-dependent correlation function following the quench from this state reads
%\bea
%&C_{ij}(t) = \frac1p\delta_{ij} + \frac{i^{j-i}}{p^2}\spm\spm \sum_{\substack{n,r,s=0\\ r\neq s}}^{p-1}\spm\spm a_{i,j}^{r,s}\spp J_{j-i+r-s}\big(2t\sin(\pi n/p)\big),
%\eea
%where
%\bea
% a_{i,j}^{r,s}=i^{r-s}\cos\spm\Big(\frac\pi2(i-j+s-r) + \frac{\pi n}{p} (i + j +r+s)\Big).
%\eea
\bea
C_{ij}(t) = C_{ij}(\infty)+\frac{i^{j-i}}{p^2}\spm\spm \sum_{n,r=1}^{p-1}\spm\spm i^{\pm r} a_{i,j}^{n,\pm r}\spp J_{j-i\pm r}\big(2t\sin(\pi n/p)\big)\spp,
\eea
where
\bea
&C_{ij}(\infty) = \frac1p\delta_{i,j} + \frac{1}{p^2}\sum_{n=1}^{p-1}(p-n)\delta_{i-j,\pm n}\spp,\\
&a_{i,j}^{n,\pm r}=\frac{\sin\spm\big(\spm\frac{n\pi r\spm}{p}\big)}{\sin\spm\big(\spm\frac{n\pi}{p}\spm\big)}\cos\spm\Big(\frac\pi2(j-i\pm r-2) - \frac{\pi n}{p} (i + j -1)\spm\Big).
 %&=\spm\sum_{s=0}^{p-1-r}\spm\spm\spm\cos\spm\Big(\frac\pi2(i-j \pm r) + \frac{\pi n}{p} (i + j +r+2s)\spm\Big),\nn\\
\eea

In the next section, we consider the initial state $|W_3\ra=\big(\frac{|\downarrow\downarrow\uparrow\ra+|\downarrow\uparrow\downarrow\ra+|\uparrow\downarrow\downarrow\ra}{\sqrt{3}}\big)\otimes\cdots$, with correlation matrix
\bea
C_{ij}(t) =C_{ij}(\infty)+\frac{2(-1)^j}{9i^{i-j+1}}\bigg[\spm\cos\Big(\frac{\pi}{6}(i+j-1)\Big)\spp J_{i-j\pm1}\big(t\sqrt{3}\big) +i\sin\Big(\frac{\pi}{6}(i+j-1)\Big)\spp J_{i-j\pm2}\big(t\sqrt{3}\big)\bigg].
\eea

\section{Time evolution of Markov gap for crystal and W-product states}

We numerically compute the time evolution of the Markov gap for several representatives of the two classes of initial states introduced above. For both classes, we find qualitative features similar to that for the state \eqref{eq:state} in the main text: a time delay that depends on the initial state and an entanglement barrier, see Figs.\,\ref{apdx:fig_plot1} and \ref{apdx:fig_plot2}. 

For the initial state $|C_{4,2}\ra=|\spm\spm\downarrow\downarrow\uparrow\uparrow\spm\cdots\ra$, which generates multiplets with four quasiparticles, we observe a clear second peak. Four quasiparticles can spread entanglement in more ways than triplets can, which may allow for the appearance of such second peak. %W-product states with $p>3$ can also display 
It is interesting to note that quenches from the state $|C_{4,2}\ra$ were recently studied through the lens of tripartite mutual information (TMI) in \cite{Caceffo:2023hns}. There it was found that the TMI is negative at intermediate times in the hydrodynamic regime. This contrasts with pairs and triplets of quasiparticles, for which the TMI vanishes, and with the ‘classically’ correlated multiplets produced by states with one fermion per unit cell (such as $|C_{p,1}\ra$ and $|W_p\ra$), for which it is nonnegative.
A negative TMI means that the mutual information is monogamous, similarly to holographic theories \cite{Hayden:2011ag}.

\medskip
\begin{figure}[h]
%\vspace{.2cm}
\centering
\includegraphics[scale=1]{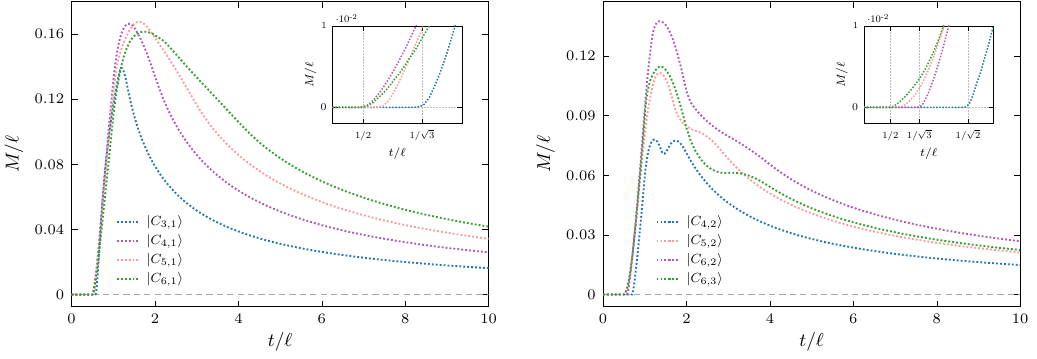}
\vspace{-5pt}
\caption{Dynamics of the Markov gap in the XX chain after quenches from $|C_{p,q}\ra$ for equal subsystem sizes $\ell_A=\ell_B=\ell=360$. The insets show close-ups around times before which the Markov gap is zero.}
\lb{apdx:fig_plot1}
\end{figure}
\begin{figure}[h]
\centering
\includegraphics[scale=1]{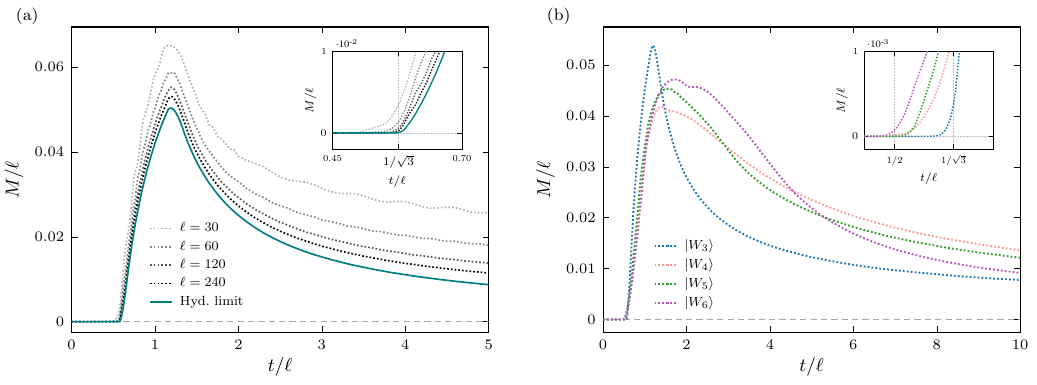}
\vspace{-5pt}
\caption{Dynamics of the Markov gap in the tight-binding chain after quenches from $|W_{p}\ra$ for equal subsystem sizes $\ell_A=\ell_B=\ell$. The insets show close-ups around times before which the Markov gap is zero. (a) Time dependence of the Markov gap following a quench from $|W_{p=3}\ra$. The dotted lines correspond to different lengths $\ell$ of the subsystems, while the solid teal line is the extrapolation to the hydrodynamic limit \mbox{$t,\ell\rightarrow\infty$} with $t/\ell$ fixed. (b) Time dependence of the Markov gap following quenches from $|W_{p}\ra$ for $p=3,4,5,6$ for fixed $\ell=180$.}
\lb{apdx:fig_plot2}
\end{figure}

\end{document}